\documentclass[runningheads]{llncs}
\usepackage[T1]{fontenc}
\usepackage{graphicx}
\usepackage{cite}
\usepackage{multirow}
\usepackage{xcolor}
\usepackage{tikz}
\usetikzlibrary{arrows,shapes,positioning,shadows,trees}
\usepackage{caption}

\newcommand\blfootnote[1]{%
  \begingroup
  \renewcommand\thefootnote{}\footnote{#1}%
  \addtocounter{footnote}{-1}%
  \endgroup
}

\begin{document}
\title{Variables are a Curse \\in Software Vulnerability Prediction}
\titlerunning{Variables are a Curse in Software Vulnerability Prediction}
%

\author{Jinghua Groppe \and Sven Groppe \and Ralf Möller}
\authorrunning{J. Groppe et al.}

%

\institute{Institute of Information Systems (IFIS), University of Lübeck, \\Ratzeburger Allee 160, 23562 Lübeck, Germany }

\maketitle              
\begin{abstract}
Deep learning-based approaches for software vulnerability prediction currently mainly rely on the original text of software code as the feature of nodes in the graph of code and thus could learn a representation that is only specific to the code text, rather than the representation that depicts the ‘intrinsic' functionality of a program hidden in the text representation. One curse that causes this problem is an infinite number of possibilities to name a variable. In order to lift the curse, in this work we introduce a new type of edge called \textit{name dependence}, a type of \textit{abstract syntax graph} based on the name dependence, and an efficient node representation method named \textit{3-property encoding} scheme. These techniques will allow us to remove the concrete variable names from code, and facilitate deep learning models to learn the functionality of software hidden in diverse code expressions. The experimental results show that the deep learning models built on these techniques outperform the ones based on existing approaches not only in the prediction of vulnerabilities but also in the memory need. The factor of memory usage reductions of our techniques can be up to the order of 30,000 in comparison to existing approaches\blfootnote{This work is part of the BMBF project with the contract number 16KIS1337.}.

\end{abstract}

\section{Introduction}\label{intro}
A number of efforts have been dedicated to applying deep learning (DL) to predict the vulnerabilities of software code. However, DL-based approaches have not achieved significant breakthroughs in this field and still have a limited capability to distinguish vulnerable code from non-vulnerable one~\cite{chakraborty2021deep}. Currently, DL approaches, both unstructured~\cite{li2018vuldeepecker, russell2018automated, dam2017automatic, zou2019mu} or structure-based~\cite{wang2016automatically, lin2018cross, pradel2018deepbugs, zhou2019devign, chakraborty2021deep}, borrowed the method used in the natural language processing to define the semantics of the full code or nodes in a code graph. The full code or a piece of the code is considered plain text like a natural language and it is first split into tokens, and each token is represented by a real-valued vector called embedding. Unstructured approaches learn the representation of the code only based on the sequence of the tokens. The sophisticated graph-based approaches learn a presentation based on the tokens appearing in each node and the relations between nodes. 

A functionality can be programmed using an infinite number of text representations and one main reason for the infinity is the arbitrariness in naming variables. For example, the functionality of the summation of two variables can be coded as $a+b$, $x1+x2$ or using any other names. Since different names have different embeddings, a DL model, which learns based on the raw code text, could only find a representation, which is specific to the code text with the used variable names, and would not be able to capture the intrinsic functionality beyond the diversity of code expression using different variable names.

We could not obtain a well-generalized model in the presence of an infinity of text code of a functionality. Therefore, we need solutions to transform an infinite number of text representations 
 of variable names into a finite number and we are suggesting such a solution in this work. Concretely, we suggest a new edge type of name dependence and an abstract syntax graph (ASG) that extends a standard abstract syntax tree (AST) with the edges of name dependence and develop a 3-property node encoding scheme based on the ASG. These techniques can be used to remove variable names from code, and greatly mitigate the semantic uncertainty of variables and the infinity of text code of a functionality. The empirical evidence presented later shows that our techniques do help DL models to learn the intrinsic functionality of the software and improve their prediction performance. 

\section{Breaking the Curse of Variables}
\label{sec:breaking} 
In order to help DL models of software vulnerability prediction to improve their generalization ability, in this section, we suggest techniques of how to transform an infinite number of text representations of varable names into a finite number. 

\subsection{Name Dependence and Abstract Syntax Graph}

\begin{figure}[t]
\captionsetup{font={small}}
\begin{minipage}{0.35\linewidth}
\centerline{\includegraphics[width=1\columnwidth]{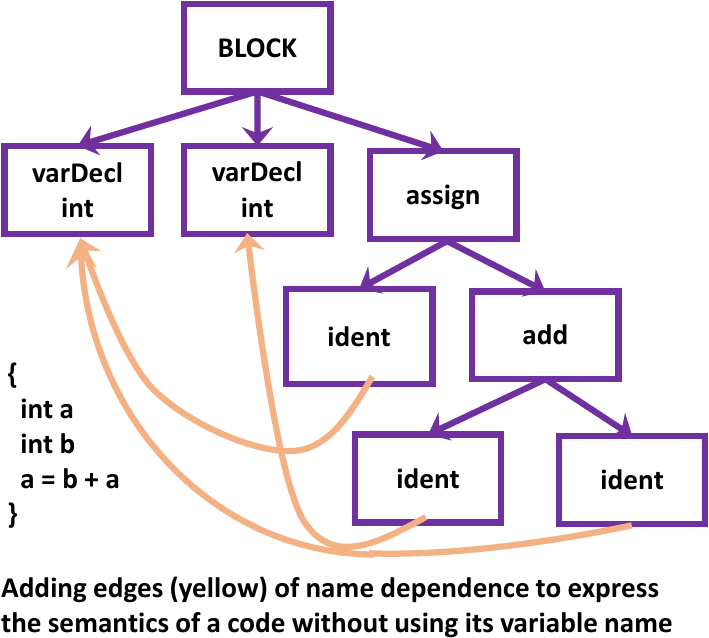}}
\caption{ASG example}
\label{fig:ASG}
\end{minipage}
\hspace{0.4cm}
\begin{minipage}{0.60\linewidth}
\captionof{table}{3-property encoding scheme} \label{tab:3-Prop}
\centering
\begin{scriptsize}
\begin{tabular}{|l||c|c|c||c|c|c|}
\hline
& \multicolumn{3}{|c||}{\textbf{3-prop. encode}}	& \multicolumn{3}{|c|}{\textbf{3-prop. encode}} \\
& \multicolumn{3}{|c||} {with variable names}	& \multicolumn{3}{|c|}{without variable} \\
\hline
\textbf{construct} & \textbf{class} & \textbf{name} & \textbf{type} & \textbf{class} & \textbf{name} & \textbf{type} \\
\hline
int a &	varDecl	& a	& int &	varDecl &	- &	int \\
If (a$\geq$0) &	control &	IF & - & control &	IF &	- \\
a*0.01	& mathOp &	mul &	-	& mathOp &	MUL &	- \\
f(a) &	call &	f	& - &	call &	f & - \\
a	& ident &	a &	int &	ident &	VAR	& int \\
stdout &	ident &	stdout	& -	& ident &	stdout	& - \\
\{…\} &	block &	- &	-	& block &	-	& - \\
10.01 &	literal &	10.01 &	float &	literal &	-	& float \\
‘Hi’	& literal &	‘Hi’ &	str	& literal &	- &	str \\
int[8] b &	varDecl	& b	& int[8]	& varDecl &	-	& int[N] \\
\hline
\end{tabular}
\end{scriptsize} 
\end{minipage}
\end{figure}

In programming languages, a variable is related to its declaration (which is either explicitly given or implied). We can determine this relation by the name of the variable. Software engineering uses the term ‘dependence’ to describe the relations between two components, like data dependence, and control dependence. To align with it, we define a new kind of dependence called \textit{name dependence} to express the relation between a variable and its declaration. In an AST with full information, the name dependence between two nodes can be inferred by the names of variables and identifiers.  When we remove the names of variables and identifiers from the AST, we lose the information on name dependence. Without the information, we will not be able to restore the semantics of the original code. 
So, we need a way to express the name dependence when names are absent. A solution is to add an edge of name dependence between two related nodes. After adding such edges, the tree structure turns into a graph structure as illustrated in Fig.~\ref{fig:ASG}, which we call \textit{abstract syntax graph} (ASG). From the graph, we can construct a fragment of code with the exact semantics as the original code, but perhaps with a different text representation, which would not be a problem at all for the task of vulnerability prediction. 

\subsection{3-Property Encoding Scheme}
Apart from the ASG, we further suggest a method to efficiently represent the nodes in a code graph, \textit{3-property encoding}, which provides a consistent description of the feature of nodes and allows DL models to infer the commons and differences between nodes easily. This 3-property encoding is developed in the context of our ASG but it can be applied to other code graphs and it is also programming languages agnostic.

In a code graph, every node represents an executable syntactic construct in code, which can be an expression, a statement or its constituent parts, like variables and constants (which are of course also executable). Currently, the piece of code that consists of the construct (with or without a notation to the construct like ‘varDecl’ and ‘add’) is used as the feature of the node. The feature is encoded by first splitting the piece of code into tokens and then averaging the embeddings of all the tokens. The code-based encoding uses the original piece of code to present the feature of a node, and at the same time, the result of encoding blurs the semantics of the original code since the averaging operation. Our 3-property encoding avoids these two issues by introducing additional information related to the language constructs.

Each language construct has its properties, which may not explicitly appear in the raw code text. Independent of specific programming languages, we found that it is enough to use three properties to describe different constructs: \textit{class}, \textit{name} and \textit{type} of data if any, and each value of the properties will be represented by a unique token. Table~\ref{tab:3-Prop} demonstrates several common language constructs and their representation with the three properties. With this property-based approach, we can encode all nodes in a consistent way, and this is a very valuable characteristic for many applications. So far, this 3-property encoding has not removed the diversity of text representations and we will further normalize this encoding scheme to mitigate the diversity as much as possible based on the name dependence and ASG.

Besides the variable names which we can remove thanks to the edges of name dependence, there are also other constructs in code, which can have any values. One of them is literals, e.g. 0.01 and ‘Hello’, which will cause similar issues as variable names, so we will also remove the concrete value of a literal. Another construct is array declarations with size, e.g. $char[8]$, $char[1024]$. We will normalize them as e.g. $char[N]$. A more refined solution could be to create several normalized data types e.g., $char[int8]$, $char[int16]$ and so on, and normalize the data type of arrays according to their sizes. For instance, any char arrays with sizes between 0 and 256 could be normalized to $char[int8]$. Table~\ref{tab:3-Prop} also provides examples of normalized representations. The definition of the classes of language constructs and the normalized tokens could vary depending on the implementation of applications and the tool for generating code graphs.

\section{Evaluation} \label{sec:evaluation}
In order to evaluate our techniques, we build four types of code graphs, AST, AST+, ASG and ASG+, for training DL models of software vulnerability prediction. AST+ is an AST extended with flow and data dependencies and control flow. ASG is an AST with the edges of name dependence and variable names removed, and ASG+ is ASG with flow and data dependencies and control flow. 

\paragraph{\textbf{Models:}} 
We develop two models (3propASG and 3propASG$+$), which use our graph structures and 3-property node encoding scheme, and two baselines (codeAST and codeAST+), which adopt the common graph structures and the pieces of code as the feature of nodes (i.e., the code-based encoding presented in Section~\ref{sec:breaking}) that is currently adopted by existing models~\cite{zhou2019devign, chakraborty2021deep}. All models share the following architecture: the input data is delivered to the layer of GGRU with one time step, the least expensive option. The output of GGRU is sent to each of three 1D convolution (Conv1d) layers with 128 filters each and perceptive fields of 1, 2, and 3 respectively, and one 1D max pooling (MaxPool1d) is applied over the output of each Conv1D to perform downsample. 
The results of the MaxPool1d layers are concatenated together and sent to the hidden layer with 128 neurons, and a 25\% dropout is applied to the output of convolution and the hidden layer. We apply Relu for non-linear transformation and the embeddings of 100 dimensions to encode tokens. 

\paragraph{\textbf{Datasets:}}
We use several real-world datasets from different open-source projects: Chromium+Debian~\cite{chakraborty2021deep}, which contains 10,699 samples and 7.05\% of which are flawed; FFmpeg+Quemu~\cite{zhou2019devign} with 13,428 samples and 43.68\% flawed; VDISC~\cite{russell2018automated} with 68,398 samples and 46.38\% flawed. The tool Joern\footnote{https://github.com/joernio/joern} is utilized to create the AST and AST+ from source code and our AST+ corresponds the code property graph of Joern.

\paragraph{\textbf{Performance:}} 
We use 80\% of the datasets as training data and 20\% for validation and evaluation. The models are trained with a batch size of 32 and a learning rate of 0.001, and the Adam optimizer~\cite{kingma2014adam} is used to minimize the loss function. Since much empirical evidence (e.g. \cite{groppe2022deep}) has shown that the pre-trained embeddings are not necessarily better than random initializations. Therefore, we use the standard normal distribution $\mathcal{N}(0, 1)$ to initialize the embeddings and train models with 10 different initializations. Table~\ref{tab:3datasets} presents the performance of models with the best F1 values. 
 The evaluation results show that the DL models based on our graph structures (ASG and ASG+) and 3-property encoding scheme outperform those based on existing graph structures (AST and AST+) and code-based encoding over all the datasets. Among these datasets, Chromium+Debian is extremely imbalanced and contains only 592 (6.92\%) samples with vulnerability. Over this dataset, our models perform significantly well with F1.
 These results are strong evidence that our techniques improve the ability of DL models to infer the functionality of code. 

\paragraph{\textbf{Memory Requirement:}} 
A huge advantage of our 3-property encoding is that it has a very low memory footprint and can process very large code graphs in comparison to the existing code-based encoding. In our experiments, an 8G memory is enough to process all the data using the 3-property encoding. In comparison, the code-based encoding requires as much as 560G memory. With the 3-property encoding, the feature of each node is represented by only three tokens. With the code-based encoding, the feature of each node is represented by a piece of raw code. Although different pieces of code will create different number of tokens and the minimal node could contain only one token, all nodes are required to have the same number of tokens. This means that all the nodes in a code graph finally consist of the maximal number of tokens.

\begin{table*}[t]
\centering
\caption{Performance of models over the datasets}
\label{tab:3datasets}
\begin{center}
\begin{scriptsize} 
\begin{tabular}{|l|c|c|c|c|c|c|c|c|}
\hline
& & & \multicolumn{2}{c|}{\tiny Chromium+Debian} & \multicolumn{2}{c|}{\tiny FFmpeg+Quemu} & \multicolumn{2}{c|}{\tiny VDISC} \\\hline
\textbf{Model} & \textbf{Graph} & \textbf{Encoding} & \textbf{Acc}	& \textbf{F1} &	  \textbf{Acc}	& \textbf{F1} &  \textbf{Acc}	& \textbf{F1} \\\hline
codeAST & AST & code & 92.01 & 30.20 & 55.36 & 57.01 & 77.82 & 75.57 \\\hline
3propASG & ASG & 3-Prop. & \textbf{92.34} &	\textbf{44.97} & \textbf{60.35} & 62.30 & \textbf{81.27} & \textbf{79.86} \\\hline
{codeAST+} & AST+ & code	 & 90.89 &	25.86 & 58.38 & 46.66 &	75.67 & 74.49 \\\hline
3propASG+ & ASG+ & 3-Prop. &	\textbf{92.34} & 44.59  & 57.04 & \textbf{62.99} & 80.94 & 79.63 \\\hline
\end{tabular}
\end{scriptsize} 
\end{center}
\vskip -0.1in
\end{table*}

\begin{table*}[!tb] 
\centering
\caption{Memory need of three samples from Chromium+Debian}
\label{tab:memory_need}
\begin{center}
\begin{scriptsize}
\begin{tabular}{|l|c|c|c|c|c|}
\hline
\multirow{2}{*}{\textbf{Hash (Code ID)}} & \multirow{2}{*}{\textbf{\#nodes}} & \multirow{2}{*}{\textbf{\#tokens}} & \multirow{2}{*}{\textbf{code-based}} &	\multirow{2}{*}{\textbf{3-prop.}} & \textbf{code-based}  \\
 &  &  &  &	 & \textbf{/3-prop.}  \\
\hline
-6552851419396579257 &	4,409	 & 33,659 &	59G	 & 5.3M	 & 11,220 \\
2388171415474875762	& 7,012 &	54,157 & 	152G & 	8.4M &	18,052 \\
5045872831385413038	& 12,077 &	96,805 & 	468G &	14.5M &	32,268 \\
\hline
\end{tabular}
\end{scriptsize}
\end{center}
\vskip -0.1in
\end{table*}

Table~\ref{tab:memory_need} provides the memory footprint required by our 3-property encoding and the existing code-based encoding for processing these samples. The comparison shows that our encoding scheme can be up to 30,000 times more efficient than the code-based encoding. This explains why existing works~\cite{zhou2019devign, chakraborty2021deep} only use the code samples with a number of nodes less than 500.

\section{Conclusions}\label{sec:conclusions}
In order to break the curse of variables, we introduce the edges of name dependence and ASG extending AST with this new type of edges and suggest a 3-property node encoding scheme based on the ASG. These techniques not only allow us to represent the semantics of code without using its variable names but also allow us to encode all nodes in a consistent way. The evaluation shows that our techniques do improve the abilities of DL models to predict software vulnerabilities. Furthermore, we also believe that the 3-property encoding will be also a useful technique for many tasks in software analysis and software engineering.


%
%
%

\bibliographystyle{splncs04}
\bibliography{references}

\begin{thebibliography}{10}
\providecommand{\url}[1]{\texttt{#1}}
\providecommand{\urlprefix}{URL }
\providecommand{\doi}[1]{https://doi.org/#1}

\bibitem{chakraborty2021deep}
Chakraborty, S., Krishna, R., Ding, Y., Ray, B.: Deep learning based
  vulnerability detection: Are we there yet. IEEE Transactions on Software
  Engineering  (2021)

\bibitem{dam2017automatic}
Dam, H.K., Tran, T., Pham, T., Ng, S.W., Grundy, J., Ghose, A.: Automatic
  feature learning for vulnerability prediction. arXiv preprint
  arXiv:1708.02368  (2017)

\bibitem{groppe2022deep}
Groppe, J., Schlichting, R., Groppe, S., M\"oller, R.: Deep learning-based
  classification of customer communications of a german utility company. In:
  Proceedings of the International Semantic Intelligence Conference. pp. 1--16.
  Springer Nature (2022)

\bibitem{kingma2014adam}
Kingma, D.P., Ba, J.: Adam: A method for stochastic optimization. arXiv
  preprint arXiv:1412.6980  (2014)

\bibitem{li2018vuldeepecker}
Li, Z., Zou, D., Xu, S., Ou, X., Jin, H., Wang, S., Deng, Z., Zhong, Y.:
  Vuldeepecker: A deep learning-based system for vulnerability detection. arXiv
  preprint arXiv:1801.01681  (2018)

\bibitem{lin2018cross}
Lin, G., Zhang, J., Luo, W., Pan, L., Xiang, Y., De~Vel, O., Montague, P.:
  Cross-project transfer representation learning for vulnerable function
  discovery. IEEE Transactions on Industrial Informatics  \textbf{14}(7),
  3289--3297 (2018)

\bibitem{pradel2018deepbugs}
Pradel, M., Sen, K.: Deepbugs: A learning approach to name-based bug detection.
  Proceedings of the ACM on Programming Languages  \textbf{2}(OOPSLA),  1--25
  (2018)

\bibitem{russell2018automated}
Russell, R., Kim, L., Hamilton, L., Lazovich, T., Harer, J., Ozdemir, O.,
  Ellingwood, P., McConley, M.: Automated vulnerability detection in source
  code using deep representation learning. In: 17th IEEE international
  conference on machine learning and applications (ICMLA). pp. 757--762 (2018)

\bibitem{wang2016automatically}
Wang, S., Liu, T., Tan, L.: Automatically learning semantic features for defect
  prediction. In: 38th International Conference on Software Engineering. pp.
  297--308 (2016)

\bibitem{zhou2019devign}
Zhou, Y., Liu, S., Siow, J., Du, X., Liu, Y.: Devign: Effective vulnerability
  identification by learning comprehensive program semantics via graph neural
  networks. Advances in neural information processing systems  \textbf{32}
  (2019)

\bibitem{zou2019mu}
Zou, D., Wang, S., Xu, S., Li, Z., Jin, H.: Vuldeepecker: A deep learning-based
  system for multiclass vulnerability detection. IEEE Transactions on
  Dependable and Secure Computing  \textbf{18}(5),  2224--2236 (2019)

\end{thebibliography}

\end{document}